\documentclass{PoS}
\usepackage{amsmath}
\usepackage{amssymb}

\title{$\bar{\Lambda}$, $\lambda_1$ and $m_b$ in three-flavor (lattice) QCD}    

\ShortTitle{$\bar{\Lambda}$, $\lambda_1$ and $m_b$ in three-flavor (lattice) QCD}

\author{\speaker{Elizabeth~D.~Freeland}\\
        School of the Art Institute of Chicago, Chicago, Illinois 60603, USA\\
        E-mail: \email{eliz@fnal.gov}}
\author{Andreas~S.~Kronfeld, James~N.~Simone, and 
   Ruth~S.~Van~de~Water\\
        Fermi National Accelerator Laboratory, Batavia, Illinois, 60510, USA}
 \author{Fermilab Lattice Collaboration with the MILC Collaboration}

\abstract{The heavy-quark expansion for inclusive semi-leptonic $B$ decays introduces $\bar{\Lambda}$ and $\lambda_1$, which are matrix elements in heavy-quark effective field theory.  We review how they can be obtained from an analysis of the heavy quark mass dependence of heavy-light meson masses in lattice QCD.  
We present preliminary results 
for the bottom quark mass, $m_b$, using $\bar{\Lambda}$ and $\lambda_1$ for the $B_s$ meson from 
a 2+1 sea-flavor unquenched calculation.}

\FullConference{XXIVth International Symposium on Lattice Field Theory\\
                July 23-28, 2006\\
                Tucson, Arizona, USA}

\begin{document}

% BACKGROUND AND OVERVIEW
\section{Background and Overview}

The study of heavy-quark physics has progressed greatly since the discovery of heavy-quark spin-flavor symmetry and the development of heavy-quark effective theory (HQET) to systematically deal with symmetry-breaking effects. 
HQET is based on a scale separation between the physics of heavy and light quarks. 
It yields expressions for observables as expansions in inverse powers of the heavy-quark mass or, alternatively, inverse powers of the heavy-meson mass. 
These expansions share a common set of HQET matrix elements which must be evaluated non-perturbatively~\cite{HQET, Falk:1992wt}.

One would like to determine these matrix elements on general principles and, more importantly,  because they are needed to ascertain the 
Cabibbo-Kobayashi-Maskawa (CKM) matrix elements via inclusive decay measurements~\cite{decay_th, Gremm:1996df}.
For example, the heavy-quark expansion for the rate of the decay $B \rightarrow X_c \,\ell \nu$ is given by~\cite{Cronin-Hennessy:2001fk}
\begin{eqnarray}
\label{decay_rate}
\Gamma = \frac{G^2_F |V_{cb}|^2M^5_B}{192\pi^3} \; 0.3689 && 
\bigg[
1 
+ 1.54 \frac{\alpha_s}{\pi} 
- 1.43\beta_0\frac{\alpha_s^2}{\pi^2} 
- 1.648\frac{ {\bar{\Lambda}}}{M_B} (1-0.87\frac{\alpha_s}{\pi}) \nonumber\\
&& \,\;\; - 0.946\frac{\bar{\Lambda}^2}{M_B^2}
- 3.185\frac{\lambda_1}{M_B^2}
+ 0.02\frac{\lambda_2}{M_B^2}
+ O(\frac{1}{M_B^3})
\bigg],
\end{eqnarray}
where $|V_{cb}|$ is the CKM matrix element of interest; $M_B$ is the $B$-meson mass; and, $\bar{\Lambda}$, $\lambda_1$, and $\lambda_2$ are scheme-dependent hadronic matrix elements defined in HQET.
Currently, the HQET matrix elements are determined by fitting measurements of various moments of heavy-meson decay distributions to corresponding HQET expressions~\cite{Cronin-Hennessy:2001fk, moments_exp, Gremm:1996df,  moments_theory}.  

In addition to using experimental measurements, one would like to calculate the HQET matrix elements from first principles.
In fact, because the same non-perturbative quantities appear in the HQET expression for the meson mass, there exists a rather direct method for calculating them using lattice QCD.  
This method was first proposed in Ref.~\cite{Kronfeld:2000gk}, where the corresponding quenched calculation was reported.

The HQET expression for the mass of a heavy-light meson is~\cite{Falk:1992wt, Gremm:1996df, meson_th} 
\begin{equation}
M = 
m 
+ \bar{\Lambda} - \frac{\lambda_1}{2m} 
- d_J \frac{z_\mathcal{B} \, \lambda_2}{2m} + O(1/m^2),
\end{equation}
where $J$ is the total meson angular momentum, and $d_0 = 3$ and $d_1 = -1$ for the pseudoscalar and vector mesons respectively.
%tracks the dependence on the heavy-quark's spin.
The mass of the heavy-light meson is $M$ and that of the heavy quark is $m$.
Working with the spin-averaged meson mass, $\overline{M}$, the equation simplifies to
\begin{equation}
\label{spin_ave}
\overline{M} -m = 
\bar{\Lambda} - \frac{\lambda_1}{2m} 
 + O(1/m^2).
\end{equation}
Since a lattice calculation allows one to work at any heavy-quark mass desired, we can generate data for a variety of heavy-light meson masses and then fit to Eq.~(\ref{spin_ave}) to determine $\bar{\Lambda}$ and $\lambda_1$.

%  HQET for Lattice QCD

The key is that the formalism of HQET applies to any underlying theory with the heavy-quark spin-flavor symmetry, such as a lattice gauge theory with heavy Wilson quarks~\cite{Kronfeld:2000ck, japanese_th}.  The lattice introduces another short distance, which can be treated perturbatively via the Wilson coefficients.  The resulting expression for the spin-averaged meson mass on the lattice is~\cite{Kronfeld:2000ck}
\begin{equation}
\label{lat_meson}
\overline{M}_1 -m_1 = 
\bar{\Lambda}(a) - \frac{\lambda_{1}(a)}{2m_2}
 + O(1/m^2).
\end{equation}
Here, the heavy-quark $a$ dependence is absorbed into the rest mass $m_1$ and the kinetic mass $m_2$, defined via the heavy-quark propagator~\cite{Mertens:1997wx}.  Discretization effects of the gluons and light quarks lead to a separate $a$ dependence of the HQET matrix elements $\bar{\Lambda}$ and $\lambda_1$.  $\overline{M}_1$ is the spin-averaged rest mass on the lattice.

% EDF version
%In using lattice QCD, one could simply discretize the continuum HQET.  That is not the formulation used here.  Instead we use descriptions of the physics obtained when HQET is applied directly to lattice QCD~\cite{Kronfeld:2000ck, japanese_th}.  This method allows one to combine Symanzik's theory of cut-off effects with the scale separation of HQET.  Heavy-quark discretization effects are contained in the HQET short-distance coefficients, and improvements can be made systematically.

%The resulting lattice expression for the spin-averaged meson mass is~\cite{Kronfeld:2000ck}
%\begin{equation}
%\label{lat_meson}
%\overline{M}_1 -m_1 = 
%\bar{\Lambda}(a) - \frac{\lambda_{1}(a)}{2m_2} 
% + O(1/m^2)
%\end{equation}
%Equation~(\ref{lat_meson}) is valid for quark masses defined in any scheme.
%The quantities $m_1$ and $m_2$ are known as the rest, $m_1$, and kinetic, $m_2$, masses~\cite{Mertens:1997wx}.  
%For the following, $m_1$, $m_2$ to refer to the masses defined by the pole in the (lattice) heavy-quark propagator~\cite{Mertens:1997wx}.
%These masses contain all the short-distance physics and, therefore, the heavy-quark discretization effects.  
%Discretization effects related to the light quarks and gluons appear as the $a$-dependence of $\bar{\Lambda}$ and $\lambda_1$.
%$\overline{M}_1$ is the spin-averaged heavy-light meson (rest) mass.

% Chiral Extrapolation
To arrive at $\bar{\Lambda}$ and $\lambda_1$ for $B^{\pm}$ and $B_d$ we would need an expression for the (spin-averaged) meson mass from chiral perturbation theory.  Continuum HQET expressions exist in the literature~\cite{Jenkins:1992hx}.  The full expression including effects from staggered quarks and HQET is being derived now. 
Because we can do simulations with light  valence quarks near or at the strange-quark mass, and because we find that the effect of sea quarks is mild, 
we obtain preliminary results 
for the bottom quark mass, $m_b$, using $\bar{\Lambda}$ and $\lambda_1$ from the $B_s$ meson.

In the following, we first discuss the lattice calculation of meson and quark masses.  We then discuss how $\overline{M}_1 - m_1$ (and hence $\bar{\Lambda}$ and $\lambda_1$) depends on sea and valence quark masses and the lattice spacing.  Finally, we calculate the binding energy for a spin-averaged $B_s$ meson and use that to arrive at a preliminary value for the bottom-quark mass.

% MESON MASSES, QUARK MASSES
\section{Meson Masses, Quark Masses: Determining $\overline{M}$, $m_1$ and $m_2$ }
% Meson Masses 
We use the MILC unquenched gauge configurations~\cite{MILC} with 2+1 flavors of sea quarks and a Symanzik-improved gluon action. We use three lattice spacings: $a = 0.18, 0.15$, and $0.12$ fm.  Both sea and light valence quarks use the ``asqtad'' staggered-fermion action~\cite{stag_fermion}.
  Light valence quarks have masses ranging from $m_q = 1.1 \,m_s$ to $0.1 m_s$, where $m_s$ is the (physical) strange quark mass.  Masses of the two light sea quarks range from approximately $0.05 \, m_s$ to $0.1 \, m_s$.
  For heavy quarks, we use the Fermilab action~\cite{El-Khadra:1996mp}.
In anticipation of the full calculation of $\bar{\Lambda}$ and $\lambda_1$,
we use seven or more heavy-quark masses at each lattice spacing.  They range in mass from heavier-than-bottom to lighter-than-charm.

Pseudoscalar and vector meson masses are obtained from two-point correlation function fits done using multi-state, constrained curve fitting~\cite{Lepage:2001ym}.  Both $\chi^2$ and fit stability are used to determine the goodness of fit.  The results are spin-averaged to obtain $\overline{M}_1$.

% Heavy-quark Masses
Equation~(\ref{lat_meson}) on its own does not specify the renormalization scheme for the masses and, hence, $\bar{\Lambda}$ and $\lambda_1$.  Although it is straightforward to obtain the pole rest and kinetic masses at the one-loop level~\cite{Mertens:1997wx, Nobes:2005dz}, 
the perturbative expansion of the pole mass is marred by infrared effects~\cite{renormalons}.  
It is, therefore, better to introduce a short-distance mass.    
Because the $\overline{\rm MS}$ mass does not run correctly 
for renormalization scales below the heavy-quark mass scale, 
%for $\mu < m$, 
it is not appropriate.
Several other short-distance mass definitions are available in the literature.  Here, we use the potential-subtracted mass, $m_{PS}$~\cite{Beneke:1998rk}, which is based on the static quark potential and  introduces a separation scale, $\mu_f$, where $\; \Lambda_{QCD} \; < \;  \mu_f \; \lesssim \;  2\; \text{GeV}$.
 
For $\alpha_s$, we use the V-scheme; scale setting, $q^*$, is done via the Brodsky-Lepage-Mackenzie (BLM) prescription~\cite{BLM}.  The value of $\alpha_s(q^*)$ is obtained from the average value of the plaquette and the four-loop $\beta$-function
as described in~\cite{Mason:2005zx}.

% SEA QUARK, VALENCE QUARK, AND LATTICE SPACING DEPENDENCIES
\section{Sea Quark, Valence Quark and Lattice Spacing Dependencies}
% Lambda-bar and lambda_1
The value of the meson binding energy, $\overline{M}_1 - m_{1,PS}$, depends upon the sea quark masses, the light valence quark masses ($m_q$) and the lattice spacing.  We will discuss each of these dependencies in turn and estimate numerically how they affect our results for the $B_s$ system.

\begin{figure}
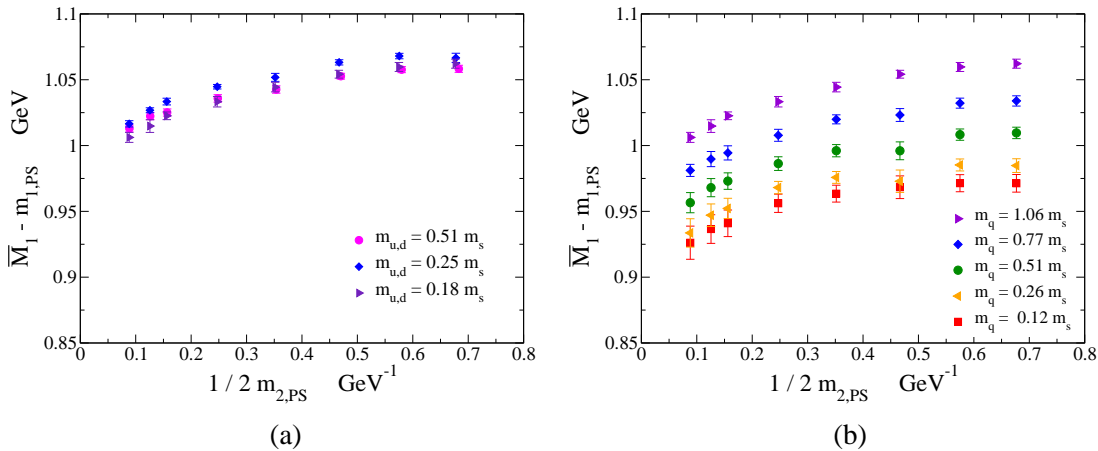

\begin{tabular}{cc}
      \includegraphics[scale=0.26]{sea_Lat06_proc}     &
      \includegraphics[scale=0.26]{valence_Lat06_proc}  \\
      $\quad$(a)  & $\quad$ (b)
\end{tabular}
\caption{(a) $\overline{M}_1 - m_{1,PS}$ vs $1/2 m_{2,PS}$ for three values of light sea quark masses; the valence quark mass $m_q = m_s$ and $a = 0.125$ fm.  (b) $\overline{M}_1 - m_{1,PS}$ vs $1/2 m_{2,PS}$ for several valence quark masses, $m_q$.  Here, $a = 0.125$ fm and $m_{u,d} = 0.18 m_s.$}   Bottom and charm quarks have values of $1/2 m_{2,PS} = 0.13$ and $0.58\, \text{GeV}^{-1}$, respectively.
\protect\label{sea_valence}
\end{figure}

We begin with the effects of the 2+1 sea quarks.  Figure~\ref{sea_valence}(a) is a plot of $\overline{M}_1 - m_{1,PS}$ vs $1/2 m_{2,PS}$ for three values of sea-quark mass ratios $m_{u,d}/m_s$.  
For each graph, $\bar{\Lambda}$ is the intercept and $\lambda_1$ is the slope, 
while the curvature is related to a combination of HQET matrix elements at
 $O(1/m^2)$~\cite{Kronfeld:2000gk}.
Figure~\ref{sea_valence}(a) allows one to view the dependence of these quantities on the sea quarks.
One can see that varying the light sea-quark mass has only a small effect on $\overline{M}_1 - m_{1,PS}$ (or $\bar{\Lambda}$ and $\lambda_1$).  
In evaluating $\overline{M}_1 - m_{1,PS}$ for $B_s$ we used, at each value of $a$, the ensemble with the lightest available $m_{u,d}$ sea quarks, and used the variation from different ensembles in our estimate of the systematic error.

Figure~\ref{sea_valence}(b)  is a plot of $\overline{M}_1 - m_{1,PS}$ vs $1/2 m_{2,PS}$ for several light valence quark masses.  
As expected, the meson mass depends strongly on the value of the light valence quark mass.
For the numerical results reported below, we will consider only mesons with a strange valence quark.  
To arrive at $\overline{M}_1 - m_{1,PS}$ for $B_s$, we use the $m_q = 1.06 \, m_s$ result and allow an uncertainty based on the value at $m_q = 0.77 \, m_s$.

\begin{figure}
 \begin{center}
         \includegraphics[scale=0.26]{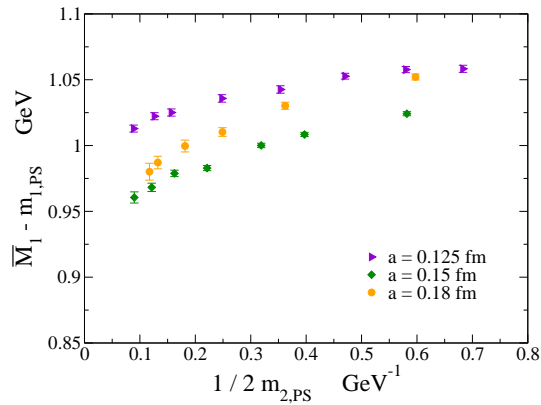}          
\end{center}
         \caption{$\overline{M}_1 - m_{1,PS}$ vs $1/2 m_{2,PS}$ for three lattice spacings: $0.125, 0.15$ and $0.18$ fm.  For all cases $ m_q = m_s $. Error bars are statistical only.
         Bottom and charm quarks have values of $1/2 m_{2,PS} = 0.13$ and $0.58\, \text{GeV}^{-1}$ respectively.}
         \protect\label{alat}
\end{figure}

Figure~\ref{alat}  is a plot of $\overline{M}_1 - m_{1,PS}$ vs $1/2 m_{2,PS}$ for three lattice spacings: $0.125, 0.15$ and $0.18$~fm.  
Results for the coarsest lattice spacing fall between the two finer spacings; the error bars shown, however, are statistical only.  Discretization errors appear from two sources.  
The first is from the light quarks and gluons.  These errors first appear at $O(\alpha_s \,a^2)$, $O(a^4)$.  
The second is from the truncation of the perturbative series for the heavy-quark masses $m_1$ and $m_2$.
Because the origins of the discretization effects from each sector are isolated, we can analyze their contributions separately.
Nevertheless, 
over the range of masses and lattice spacings we are working with,
the two uncertainties are of comparable size, so
 the behavior of the total discretization error is non-trivial.  
It is expected that the inclusion of two-loop effects would clarify the results shown here. 
To estimate the error due to truncation, we take the 2-loop contribution to be a factor of $\alpha$ smaller than that of the 1-loop contribution. 
Results from extrapolations of $\overline{M}_1 - m_{1,PS}$ in $a^2$ and $a^4$ were used to assign a systematic error for the discretization of light quarks and gluons.
Our final error budget includes an uncertainty in the determination of $a$ itself.
%To obtain the binding energy of the (spin-averaged) $B_s$ meson, we average the results at the $a = 0.125$ fm and $0.15 $ fm spacings.

% RESULTS
\section{Result for the $b$-quark mass}
Focusing on $m_b$, we calculate the binding energy of a (spin-averaged) $B_s$ meson, 
by averaging the results at the $a = 0.125$ fm and $0.15 $ fm spacings.
%given by $\overline{M}_1 - m_{1,PS}$.  
Our result is   $0.99(18)$~GeV in the potential-subtracted scheme with a factorization scale of $\mu_f = 2.0$~GeV.
Using this value of the binding energy, we can make a preliminary estimate of the value of the bottom-quark mass.
\begin{equation}
m_b = M^{\rm exp} - (\overline{M}_1 - m_{1,PS})
\end{equation}
where $M^{\rm exp}$ is the experimentally measured, and spin-averaged, value of the $B_s$ mass; we use the value of $5.402(2)$~GeV from values from the Particle Data Group~\cite{PDG06}. This yields  a preliminary value 
$m_{b,PS}=4.41(18)$~GeV.
%\begin{equation}
%m_{b,PS}=4.41(18) \; \text{GeV}.
%\end{equation}
  We do not quote a result in the $\overline{\rm MS}$ scheme at this time.  For comparison, a QCD sum rule calculation~\cite{Pineda:2006gx} obtains,  $m_{b,PS} = 4.52(6)$~GeV at $\mu_f = 2.0$~GeV and $m_{b,\overline{\rm MS}} = 4.19(6)$~GeV.

Table~\ref{error_budget} provides a preliminary budget of the uncertainties in this calculation.  The two largest are the uncertainty due to truncation of the QCD perturbation theory for the quark masses and the uncertainty due to the continuum extrapolation.  All uncertainties are added in quadrature to arrive at the total.
\begin{table}
\begin{center}
\begin{tabular}{ll}
\hline \hline
{\bf Source}    &       {\bf GeV} \\ \hline
statistical                    & 0.005 \\ 
inputs ($a$, $\kappa_b$, $\kappa_{\rm crit}$, $u_0$)     & 0.041 \\
sea-quark mass dependence        & 0.04 \\ 
strange-quark mass tuning    & 0.025 \\ 
perturbation theory (heavy quark discretization)         & 0.10 \\ 
light quark and gluon discretization    & 0.14 \\  \hline 
{\bf total }             &    {\bf 0.18}  \\
\hline \hline
\end{tabular}
\caption{Uncertainties in the quantity $\overline{M}_1 - m_{1,PS}$  for a spin-averaged $B_s$ meson.}
\label{error_budget}
\end{center}
\end{table}

% SUMMARY AND OUTLOOK
\section{Summary and Outlook}

We report a preliminary calculation of the bottom quark mass using a (lattice) HQET calculation of the spin-averaged $B_s$ binding energy.  We use lattice QCD with 2+1 flavors of staggered sea quarks.  Heavy-light mesons are constructed from a staggered valence and  Fermilab heavy quark.
We find $m_{b,PS}=4.41(18)$~GeV in the potential-subtracted scheme with a factorization scale of $\mu_f = 2.0$~GeV.  The dominant uncertainties in this calculation can be reduced by the inclusion of 2-loop effects in the perturbative expansions for $m_1$ and $m_2$, and with improved understanding of
light quark and gluon discretization effects.
Future work will include the calculation of  HQET matrix elements $\bar{\Lambda}$ and $\lambda_1$ for $B^{\pm}$ and $B_d$, which can be used in the determination of CKM matrix elements from their inclusive, semileptonic decays.

% ACKNOWLEDGEMENTS
\section*{Acknowledgments}
E.D.F. would like to thank the Fermilab Theoretical Physics Department for their hospitality while this work was being done, and Don Holmgren and Amitoj Singh for their help with computing issues.
%We would also like to thank Tetsuya Onogi for useful comments made during the conference.
E.D.F. is supported in part by the American Association of University Women.  Fermilab is operated by Universities Research Association Inc., under contract with the DOE.

\end{document}